\documentclass[
twocolappendix,
twocolumn,
trackchanges
]{aastex631}

\usepackage{CJKutf8}
\usepackage{hyperref}
\usepackage{amsmath}
\usepackage{physics}
\usepackage{amssymb}
\usepackage{enumerate}
\usepackage{bm}
\usepackage{xspace}
\usepackage{times}
\usepackage{pifont}

\newcommand{\eg}{e.g.,~}
\newcommand{\ie}{i.e.,~}

\newcommand{\orcid}[1]{\href{https://orcid.org/#1}{
\includegraphics[width=10pt]{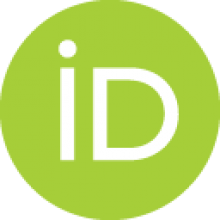}}}
\newcommand{\BHAC}{\texttt{BHAC+}\xspace}
\newcommand{\FUKA}{\texttt{FUKA}\xspace}

\newcommand{\FIL}{\texttt{FIL}\xspace}
\newcommand{\HO}{\texttt{hand-off}\xspace}

\begin{document}

\title{The initial spin matters: the impact of rapid rotation on
  magnetic-field amplification at merger}

\author{Harry Ho-Yin Ng\:\orcid{0000-0003-3453-7394}}
\affiliation{Institut f\"ur Theoretische Physik, Goethe Universit\"at,
  Max-von-Laue-Str. 1, 60438 Frankfurt am Main, Germany}

\author{Jin-Liang Jiang\:\orcid{0000-0002-9078-7825}}
\affiliation{Institut f\"ur Theoretische Physik, Goethe Universit\"at,
  Max-von-Laue-Str. 1, 60438 Frankfurt am Main, Germany}
\affiliation{Department of Astronomy, School of Physics and Astronomy,
  Key Laboratory of Astroparticle Physics of Yunnan Province, Yunnan
  University, Kunming 650091, People's Republic of China}

\author{Luciano Rezzolla\:\orcid{0000-0002-1330-7103}}
\affiliation{Institut f\"ur Theoretische Physik, Goethe Universit\"at,
  Max-von-Laue-Str. 1, 60438 Frankfurt am Main, Germany}
\affiliation{School of Mathematics, Trinity College, Dublin 2, Ireland}
\affiliation{CERN, Theoretical Physics Department, 1211 Geneva 23,
  Switzerland}

\date{\today}

\begin{abstract}
A couple of milliseconds after the merger of a binary system of neutron
stars can play a fundamental role in amplifying the comparatively low
initial magnetic fields into magnetar strengths. The basic mechanism
responsible for this amplification is the Kelvin-Helmholtz instability
(KHI) and we here report the first systematic study of the impact of
rapid rotation on the KHI-amplification process exploiting
general-relativistic magnetohydrodynamic simulations at very
high-resolutions of $35\,{\rm m}$. Concentrating on four different
spinning configurations, we find that aligned, anti-aligned, and mixed
(aligned/anti-aligned) spin configurations lead to markedly different
growth rates of the electromagnetic (EM) energy, field topologies, and
vortex properties when compared to the irrotational case. These
differences arise from intrinsic variations in the system dynamics, such
as tidal deformation, collision strength, and contact surface area, with
the anti-aligned configuration producing the largest vorticity and growth
in EM energy. Importantly, while different spin configurations lead to
significantly different initial growth rates of the poloidal/toroidal
components, all systems converge to a specific topological partition. Our
simulations are confined to a short window in time, but the different EM
energies produced as a result of spin will imprint the EM emission at
merger and provide information on the spinning state at merger.
\end{abstract}

\section{Introduction}
\label{sec:intro}
Binary neutron star (BNS) mergers are rich multi-messenger events that
emits gravitational waves (GWs)~\citep{Baiotti2016, Paschalidis2016,
  Radice2020b}, and a large frequency range of electromagnetic (EM)
signals, originated from magnetized flaring~\citep{Most2023,
  Musolino2024b, Jiang2025}, short gamma-ray burst (sGRB) due to the
launching of relativistic jets~\citep{Giacomazzo2011b, Rezzolla:2011,
  Baiotti2016, Ciolfi2020, Kiuchi2023, Gottlieb:2023a, Gottlieb2024,
  Chen2025b}, afterglow~\citep{Lyman2018, Hajela2019, Hajela2022_ag} and
kilonova powered by the radioactive decay of $r$-process synthesised
ejecta~\citep{Radice2016, Metzger2017, Bovard2017, Zhu2021b, Combi2022,
  Fujibayashi2023, Kawaguchi2023, Cheong2024c, Ng2024c}.

The first multi-messenger BNS merger observed in 2017 provided both GW
and EM detections~\citep{Abbott2017, Abbott2017_etal, FermiLat2017,
  Hajela2019}. Due to the highly nonlinear dynamics of BNS mergers,
intensive modelling with general-relativistic magnetohydrodynamics
(GRMHD) simulations is required. A particularly challenging aspect is the
accurate treatment of matter-magnetic field interactions. While these
interactions are not expected to play a role in the
inspiral~\citep{Giacomazzo:2009mp, Zhu2020}, they are crucial at merger
and afterwards. Yet, despite significant progress in modelling magnetized
BNS mergers \cite[see, \eg][]{Moesta2020, Combi2023, Most2025,
  Fields2025, Celora2025}, key MHD instabilities -- such as the
Kelvin-Helmholtz instability (KHI) and the magnetorotational instability
(MRI) -- remain difficult to fully resolve because of their strong
sensitivity to grid resolution, making them computationally expensive to
capture~\citep[see, \eg][]{Siegel2013, Palenzuela_2022PRD, Kiuchi2024}.

In particular, at merger the KHI can amplify the comparatively weak
magnetic fields of the old and cold pre-merger neutron stars (NSs) by
orders of magnitude and over a timescale that is of a couple of
milliseconds only~\citep{Price06, Giacomazzo:2014a, Kiuchi2015a,
  Chabanov2022, Kiuchi2023, Aguilera-Miret2025, Gutierrez2025,
  Neuweiler2025b}. However, several issues regarding KHI-amplified
magnetic fields remain unclear: (i) a saturated growth has not yet been
achieved, except in the large eddy simulations (LES) employing
phenomenological subgrid models~\citep{Aguilera-Miret2024,
  Aguilera-Miret2025}; (ii) the influence of the initial magnetic-field
topology and magnetic field strength~\citep{Chabanov2022, Jiang2025,
  Gutierrez2025, Cook2025}; (iii) the impact of the initial conditions in
terms of the masses, mass ratio, NS spins, and equation of state
(EOS)~\citep{Dudi2021, Bamber2024, Musolino2024b, Neuweiler2025b}.

While direct simulations that have so far represented the most robust way
to explore these issues, progress on any of them has been hindered by the
enormous computational costs. We here address in part point (iii) by
exploiting a novel ``hybrid'' approach which reduces significantly the
computational costs by making use of two GRMHD codes optimized for
different phases (inspiral/merger) of BNSs~\citep{Jiang2025}. In this
way, we are able to use very high resolutions of $35\,{\rm m}$ on rather
large portions of the domain and, at the same time, consider a coarse but
broad sampling of the space of parameters. In this way, we can assess how
the initial spin affects both the amplification process and the
properties of the turbulence generated during the merger and early
post-merger phases.

\section{Numerical setup}
\label{sec:num_setup}

\begin{table}[]
  \centering
  \footnotesize
  \begin{tabular}{lccccc}
    \hline 
    \hline 
    binaries & $t_{\rm mer}$ &  $t_{\rm sh}-t_{\rm mer}$ & $t_{\rm sh, 1/2}-t_{\rm mer}$ & $t_{_{\rm KHI}}$  
    & $v_{\rm a}$\\
    \phantom{binaries} & $[\rm ms]$ &  $[\rm ms]$ & $[\rm ms]$ & $[\rm ms]$  
    & $[c]$\\
    \hline
    \texttt{IR}        & $ 51.6$  & $-0.06$   & $0.63$   & $0.69$    &    $0.013$  \\
    \texttt{UU}        & $ 57.0$  & $-0.01$   & $1.34$   & $1.35$    &    $0.010$  \\
    \texttt{DD}        & $ 34.4$  & $-0.32$   & $0.69$   & $1.01$    &    $0.040$  \\
    \texttt{DU}        & $ 44.6$  & $-0.49$   & $0.52$   & $1.01$    &    $0.026$  \\
    \hline
    \hline 
  \end{tabular}
  \caption{Summary of the relevant times for the four binaries simulated:
    irrotational \texttt{IR}, aligned \texttt{UU}, anti-aligned
    \texttt{DD}, or mixed \texttt{DU}. Reported are: the merger time
    $t_{\rm mer}$, the ``shear time'' $t_{\rm sh}$ when the KHI develops
    robustly (\ie when the growth rate $\gamma$ is $\gamma = \gamma_{\rm
      max}/3$, where $\gamma_{\rm max}$ is the maximum growth rate), the
    ``half-time'' $t_{\rm sh, 1/2}$ when the KHI can be considered
    globally quenched (\ie when $\gamma = \gamma_{\rm max}/2$ and
    $\gamma_{\rm max}$ have both been reached), the KHI duration time
    $t_{_{\rm KHI}} := t_{\rm sh, 1/2}-t_{\rm sh}$, and the approaching
    velocity $v_a$ of the maximum rest-mass density when the two NSs have
    a separation of $18\,M_{\odot}$.}
  \label{tab:ID}
\end{table}

The simulations reported below are obtained by applying \HO procedure
developed in~\cite{Ng2024b}, and employed in~\cite{Jiang2025}, which
focus on the BNS postmerger stage. In essence, a \HO procedure consists
of the exchange of data between two GRMHD codes, \FIL~\citep{Most2019b,
  Most2019c}, which solves the full set of Einstein equations with the
constraint-damping formulation of the Z4
formulation~\citep{Bernuzzi:2009ex, Alic:2011a} and \BHAC~\citep{Ng2024b,
  Jiang2025}, which solves the Einstein equations under the conformal
flatness condition (CFC)~\citep{Cordero2009, Bucciantini2011, Cheong2021}
together with a new gravitational-wave radiation-reaction (GWRR)
treatment~\citep{Jiang2025} improving previous
formulations~\citep{Faye2003, Oechslin07a}. Both codes employ a
fourth-order accurate conservative finite-difference scheme with a WENO-Z
reconstruction method~\citep{Acker2016, Most2019b} coupled with an HLL
Riemann solver~\citep{Harten83, Rezzolla_book:2013}. While \FIL employs a
vector-potential based constrained transport scheme~\citep{Etienne2015},
\BHAC uses a magnetic-field based upwind constrained
transport~\citep{Olivares2019}. Both codes use a third-order Runge-Kutta
method for the time integrator with a Courant-Friedrichs-Lewy (CFL)
factor of $\mathcal{C}_{\rm CFL} = 0.2$ and Cartesian coordinates with a
reflection symmetry on the $z=0$ plane; it was recently shown that this
symmetry is accurately preserved in the first ms after
merger~\citep{Gutierrez2025}.

To explore the impact of the stellar spin on the magnetic-field
amplification by KHI, we construct equal-mass rapidly spinning BNS
configurations having the same total mass of $2.584\,M_{\odot}$, initial
separation of $38\,M_{\odot}$, and being governed by
temperature-dependent TNTYST EOS~\citep{Togashi2017}. The initial
binaries are in quasi-circular equilibrium and computed with the
open-source code \FUKA~\citep{Papenfort2021, Tootle2021} and each star
has a dimensionless spin $\chi := J_{\rm s}/M^2 = 0.35$, where $M$ and
$J_{\rm s}$ are respectively the gravitational mass and spin angular
momentum, with the latter being parallel to the orbital angular momentum
$J$. For our EOS, this corresponds to a rotation frequency of $620\,{\rm
  Hz}$, which is smaller than the largest observed frequency of
$716\,{\rm Hz}$ ($\chi \sim 0.4$) for
PSR~J1748--2446ad~\citep{Hessels2006}. To cover the possible space of
parameters, we consider four configurations that are distinguished by the
spin orientation, \ie binaries that are either: irrotational \texttt{IR},
aligned \texttt{UU}, anti-aligned \texttt{DD}, or mixed \texttt{DU} (see
Table.~\ref{tab:ID} for a summary)

Because of different mesh-refinement strategies [moving boxes in
  \FIL~\cite{Schnetter-etal-03b} and quadtree--octree blocks in
  \BHAC~\citep{Porth2017, Keppens2021}], the grid setup is different in
the two codes. More specifically, the \FUKA initial data is first evolved
by \FIL, with the outer boundary set at $1000\,M_{\odot}$ and employing
six refinement levels. The finest refinement level encompasses a box
region of $[-16, 16]\,M_{\odot}$ ($[-23.6, 23.6]\,{\rm km}$) in each
direction with a finest resolution of $300\,{\rm m}$. In all simulations, a weak
and purely poloidal magnetic field is introduced in each star slightly
before the merger (\ie at time $t- t_{\rm mer} \simeq - 1.2\,{\rm ms}$)
and its strength is adjusted such that all stars have the same maximum
magnetic-field strength of $|B|_{\rm max} \simeq 3.8\times10^{11}\,{\rm
  G}$ in the Eulerian frame [see coefficient $A_b$ in Eq.~(15)
  in~\cite{Jiang2025}]; in all cases, the magnetic-dipole vector and the
spin angular-momentum vectors are parallel (see also~\cite{Kawamura2016,
  Ruiz2018, Ruiz2020} for previous work with different magnetic-field
orientations). Because stars magnetized in this way do not correspond to
consistent solutions of the GRMHD system~\citep[see, \eg][for examples of
  consistent solutions]{Bonazzola1996, Frieben2012, Cheong2024d}, the
``late'' magnetization of the star is made to avoid artificial
amplification/dissipation of magnetic fields due to inconsistencies
between initial data and magnetic fields.

The large initial separation of $\sim 56\,{\rm km}$ -- necessary to
obtain convergence when considering such high initial spins -- also leads
to a large number of inspiral orbits before merger (\ie between 9 for the
\texttt{DD} and 15 for the \texttt{UU} configurations) that are computed
by the \FIL code. Right before the formation of the shearing layer and
the occurrence of KHI ($t-t_{\rm mer} \simeq -1\,{\rm ms}$), the \HO to
\BHAC takes place, with a change in the outer boundary from
$2000\,M_{\odot}$ (in \FIL) to $100\,M_{\odot}$ (in \BHAC) and an
increase to eight refinement levels, thus having the finest resolution of
$35.0\,{\rm m}$. We note that even higher resolutions of $12.5\,{\rm m}$
have been used in the recent past~\citep[see, \eg][]{Kiuchi2023} on
volumes smaller than those we covered at $35\,{\rm m}$ corresponding to a
cubic box of extent $[-12,12]\,\mathrm{km}$. Our experience with very
high resolutions restricted to small volumes has revealed that this can
lead to artifacts that can easily go unnoticed. To avoid them, and keep
the computational costs comparable, we have decreased the resolution to
$35\,{\rm m}$ but increased the volume covered.

Obviously, the \HO implies a change in the way the Einstein equations are
solved and comes with a number of advantages and one (minor)
disadvantage. First, the CFC scheme(s) has been used with success to
study core-collapse supernovae~\citep{Dimmelmeier02a, Ott07b}, isolated
NSs~\citep{Yoshida2012b, Ng2021, Yip2025}, and even BNS
mergers~\citep{Oechslin07a, Bauswein2015}. More
recently,~\cite{Jiang2025} have also shown that an improved CFC scheme --
the extended CFC (xCFC) complemented with GWRR terms -- recovers with
high accuracy a full general-relativistic description of the remnant of a
binary merger. Second, the timestep can be set to be much larger than in
a full numerical-relativity code since it is no longer limited by the
speed of light but by the speed of sound. Third, the solution of the set
of elliptic xCFC constraint equations does not need to be computed at every
timestep but at a fraction of the GRMHD timesteps. When combined, these
advantages lead \BHAC to have a net saving of $\sim 80\%$ of the
computational costs to be sustained by \FIL~\citep{Jiang2025}. At the
same time, the basic xCFC scheme does not account for energy and momentum
losses via GWs. These losses can be included with the GWRR corrections,
which then yield evolutions that reproduce very well the bulk dynamics of
full general-relativistic simulations. Furthermore, these losses are not
the main source of error, which remains in the finite resolution of the
simulations (the relative difference between the xCFC solution and the
full numerical-relativity solution is $\sim 3\%$--$12\%$; see
appendix~\ref{sec:app_B} for more details).

In summary, we find that the substantial reduction in computational cost
and the ability to carry out a broad investigation of the space of
parameters employing very high resolutions that would be otherwise
unsustainable, shadows the relatively small errors introduced by the
approximate gravitational description.

\section{Results}
\label{sec:results}

\subsection{Spins, KHI and Shocks}

Being an instability, the KHI will be characterised by an exponential
behaviour in time, so that, dynamical quantities such as the total EM
energy $E_{_{\rm EM}}$ [see Eq.~(A3) in~\cite{Jiang2025} for a
  definition] will have a dynamical behaviour of the type $E_{_{\rm EM}}
\sim (E_{_{\rm EM}})|_{t=t_{\rm sh}} \exp (\gamma\,t)$, where $\gamma :=
\dot{E}_{_{\rm EM}}/E_{_{\rm EM}}$ is the growth rate and $t_{\rm sh}$ is
the ``shear time'', \ie the time when the exponential behaviour is first
recorded. In a linear regime and under idealised conditions, $\gamma$ is
assumed to be a constant; however, in the highly nonlinear conditions in
which the KHI develops at a BNS merger, the growth rate is actually time
dependent, $\gamma=\gamma(t)$, and it is therefore useful to introduce
$t_{\rm sh}$ and $t_{\rm sh, 1/2}$ respectively as the times when
$\gamma$ reaches either $1/3$ or $1/2$ of its maximum value on either
side of it [\ie $t_{\rm sh}$ ($t_{\rm sh, 1/2}$) marks the time when
  $\gamma = \gamma_{\rm max}/3$ ($\gamma = \gamma_{\rm max}/2$) before
  (after) the maximum value $\gamma_{\rm max}$ is reached].  Within this
notation, $t_{\rm sh}$ and $t_{\rm sh, 1/2}$ can be taken to define the
effective times when the KHI robustly develops and is globally quenched,
respectively. We remark, however, that these serve mostly as reference
times among the various spinning configurations and it is possible that
the KHI is locally active before $t_{\rm sh}$ and after $t_{\rm sh,
  1/2}$. These two times are useful to compare the growth rates across
the various cases considered and to determine a window in time during
which the KHI is robustly active. In the presence of spin, in fact, an
artificial linear growth is measured even before the NSs enter in
contact, hence the need for $t_{\rm sh}$ for approximately indicating the
starting moment of exponential growth and the formation of shear layer.
The magnetic field is still growing rapidly because of the
turbulent amplification following the KHI, hence the need for introducing
$t_{\rm sh, 1/2}$~\citep[see also][for a more extended
  discussion]{Chabanov2022}. The values of $t_{\rm sh}$ and $t_{\rm sh,
  1/2}$ are reported in Tab.~\ref{tab:ID}.

\begin{figure}[t!]
  \includegraphics[width=0.47\textwidth]{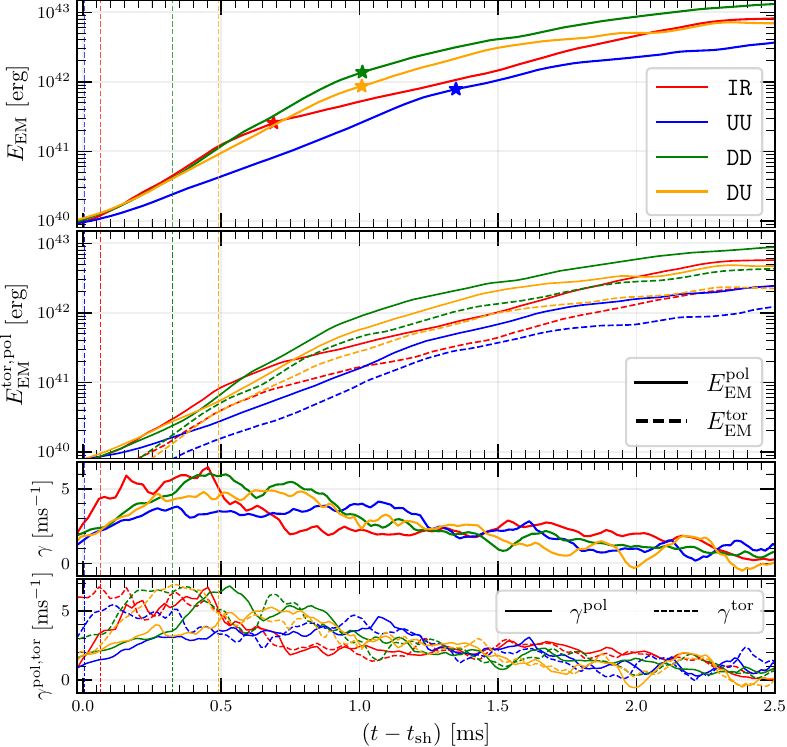}
  \caption{\textit{Top panel:} evolution of the total EM for the four
    binaries considered, star symbols indicating when the KHI can be
    considered to be globally quenched ($t = t_{\rm sh,
      1/2}$). \textit{Second panel:} evolutions of the EM energy in the
    poloidal (solid lines) and toroidal components (dashed
    lines). \textit{Third panel:} Evolutions of the KHI growth rate
    $\gamma := \dot{E}_{_{\rm EM}}/E_{_{\rm EM}}$. \textit{Bottom panel:}
    evolutions of the KHI growth rates in the poloidal (solid lines) and
    toroidal components (dashed lines). In all panels, the vertical
    dashed lines mark the merger time $t_{\rm mer}$ for each binary.}
  \label{fig:fig1}
\end{figure}

Figure~\ref{fig:fig1} shows, relative to $t_{\rm sh}$ and for the four
binaries considered here, the evolutions of the total EM energy $E_{_{\rm
    EM}}$ (top panel), of its poloidal and toroidal components $E^{\rm
  pol}_{_{\rm EM}}$, $E^{\rm tor}_{_{\rm EM}}$ [second panel from the
  top; see Eqs.~(A3)--(A5) in~\cite{Jiang2025} for a definition], the
evolution of the total growth rate $\gamma$ (third panel), and of the
various components (bottom panel); all energies are computed over the
whole computational domain for all configurations in Tab.~\ref{tab:ID}.

\begin{figure*}[ht!]
  \includegraphics[width=1.0\textwidth]{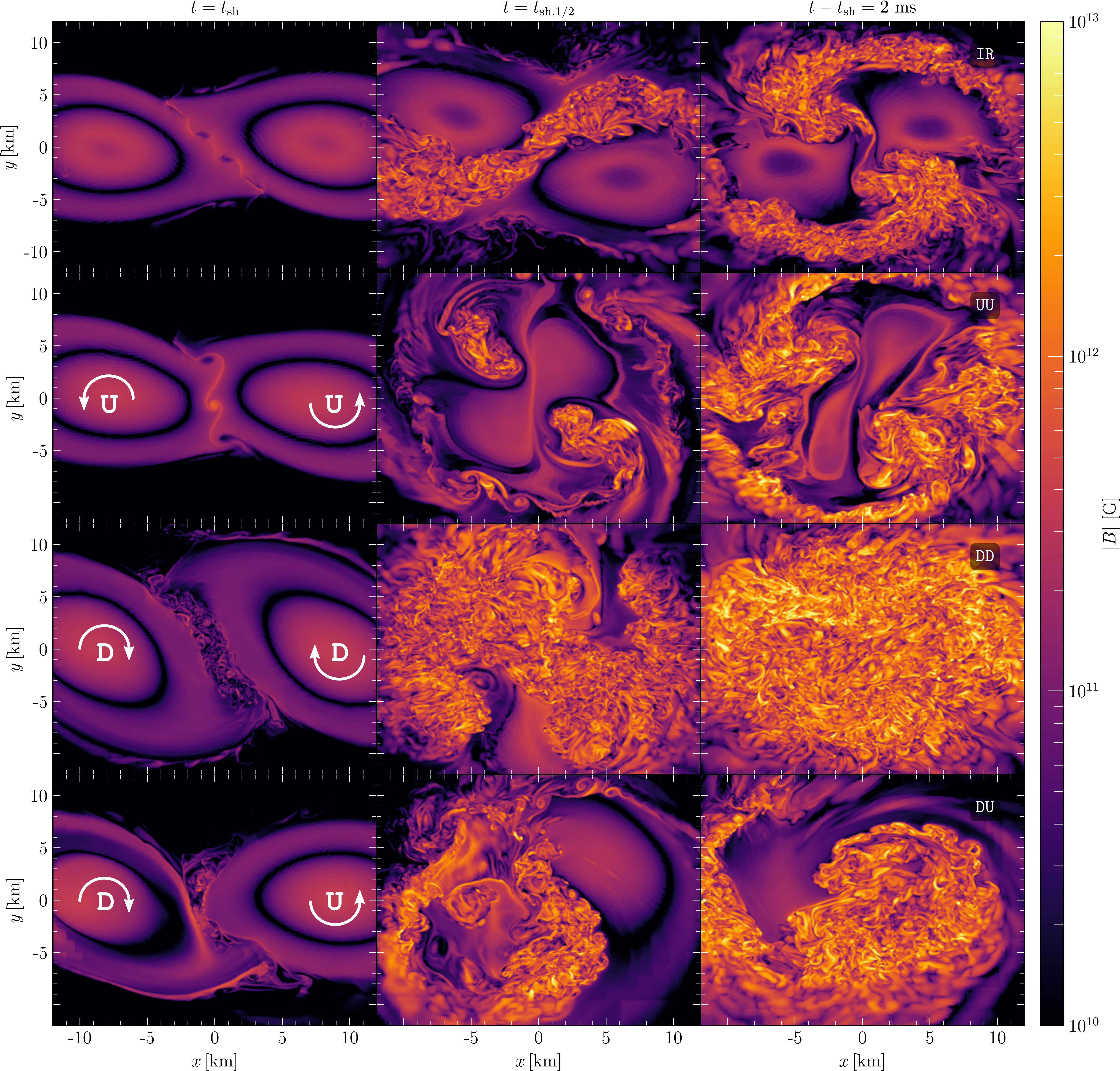}
  \caption{Distributions of the magnetic-field strength $|B|$
    on $z=0$-plane for all cases considered (different rows) at times $t
    = t_{\rm sh}$, $t_{\rm sh, 1/2}$, and $t-t_{\rm sh}=2\,{\rm ms}$
    (different columns). Note that all figures have been rotated so that
    the maxima of the rest-mass density of the two NSs at $t = t_{\rm
      sh}$ are on the $y=0$ line (for the \texttt{DU} binary, only the
    density maximum of aligned-spinning star is on $y=0$ line.). Each
    panel reports data covered by the finest resolution box with $\Delta
    x \simeq 35\,{\rm m}$. The arrows in first column indicate the
    spin-alignment of each star.}
  \label{fig:fig2}
\end{figure*}

A number of considerations can be made after analysing
Fig.~\ref{fig:fig1}. The first, and possibly most important in our study,
is that the \textit{``initial spin matters''}, namely, the KHI-induced
amplification of the magnetic field and the subsequent turbulent
amplification are strongly influenced by the spin of merging stars. This
can be appreciated from the top panel of Fig.~\ref{fig:fig1}, which shows
how the binary with aligned spins, \ie the \texttt{UU} binary, has the
lowest EM magnetic energy at the end of the time interval considered and
among all cases considered. By contrast, the binary with anti-aligned
spins, \ie the \texttt{DD} binary, has the highest EM energy, which is
$\sim 5$ times larger than for the \texttt{UU} binary. Interestingly, the
mixed-spin binary, \ie the binary \texttt{DU}, has a final EM energy that
is comparable with that of the irrotational binary \texttt{IR} even
though the evolution is rather different.

Second, note the complex relation between the merging time $t_{\rm mer}$
and the time at which the KHI starts developing robustly, $t_{\rm
  sh}$. While the \texttt{DD} binary is the first to merge and the
\texttt{UU} binary the last one\footnote{The delay in the merger time
with increasing spin is also known as the ``hang-up'' effect and reflects
the spin-orbit angular-momentum interaction appearing already at the
1.5~post-Newtonian (PN) order~\citep[see, \eg][]{Blanchet2014}.}, the
binaries \texttt{DU} and \texttt{IR} follow in this order (see the dashed
vertical lines in the top panel, which mark $t_{\rm mer}$ and
Tab.~\ref{tab:ID}). These differences in the times of shear emerge
because the exponential increases in EM energy occur earlier for the two
binaries with an anti-aligned NS (\ie with an earlier $t_{\rm sh}$), as
these stars are more strongly tidally deformed during the late-inspiral
phase due to larger spin-dependent dynamical tidal effects and
counter-rotation of the down-spin star relative to the orbital
motion~\citep[see][for details]{Steinhoff2021}.

Third, the poloidal component of the EM energy (solid lines in the second
panel from the top) is about twice that of the toroidal component (dashed
lines) and this ratio is very similar for all cases after $t_{\rm sh,
  1/2}$, which are marked with coloured stars (see also below for a
detailed discussion).

Fourth, the growth rates of the KHI are also rather different in the four
cases and during the course of the evolution. Interestingly, the
irrotational \texttt{IR} binary has the largest initial growth rate (see
third panel from the top), while the \texttt{UU} binary has the smallest
one. At late times, \ie for $t > t_{\rm sh, 1/2}$ and when the KHI has
been quenched, the growth rates are remarkably similar in all four
binaries.

Fifth, after the KHI stage ends and before the winding and MRI phases,
all cases enter a turbulent amplification stage~\citep{Siegel2013}. This
stage represents nonlinear amplification driven by interactions among
KHI-induced small-scale eddies and the reorganisation of the fluid prior
to reaching a quasi-equilibrium state. Although the growth rate during
this stage is lower than in the KHI phase, it helps counteract EM energy
attenuation caused by strong pressure gradients resulting from core
contraction and expansion, mass ejection, and numerical resistivity.
Finally, and not surprisingly, the initial growth rates of the toroidal
components are the largest but not much larger than those of the poloidal
ones (see fourth panel from the top).

In order to understand why and how the different spin configurations
influence the magnetic-field amplification via the KHI, it is useful to
study the equatorial distribution of the magnetic field strength $|B| :=
\sqrt{B^i B_i}$ and the associated vortical structure. This is reported
in Fig.~\ref{fig:fig2} for all binaries (different rows) and at times
$t_{\rm sh}$, $t_{\rm sh, 1/2}$, and $t-t_{\rm sh}=2\,{\rm ms}$
(different columns). For a more detailed comparison of the properties of
the KHI-unstable shear layer when the two stars come into contact, we
have rotated the distributions for the \texttt{IR}, \texttt{UU}, and
\texttt{DD} cases so that the maxima of the rest-mass density $\rho$ of
the two NSs at $t = t_{\rm sh}$ (first column) are on the $y=0$ line. On
the other hand, for the \texttt{DU} binary, where one component is
tidally disrupted (see below), the rotation is made such that only the
rest-mass density maximum of the star with aligned spin is on the $y=0$
line.

The different columns in Fig.~\ref{fig:fig2} are chosen so as to contrast
the evolutions at three important times. More specifically, the first
column, at $t = t_{\rm sh}$, illustrates the stellar shapes at the onset
of the shear layer and vortices. The second column, at $t_{\rm sh, 1/2}$,
marks the end of the KHI, as that this time the normalized growth rate
drops to half of its peak value. At this stage, the KHI-driven eddies
have expanded and been advected into wider regions of the remnant, and in
some configurations they even penetrate into the deeper stellar
interior. The third column, at $t - t_{\rm sh} = 2\,{\rm ms}$, represents
a time that is close to the end of the simulated window, where the KHI
stage has concluded and the subsequent turbulent amplification has begun,
but before the winding and MRI stages. We will next discuss in detail the
evolution in these three stages.

\begin{figure*}[t!]
  \includegraphics[width=1.0\textwidth]{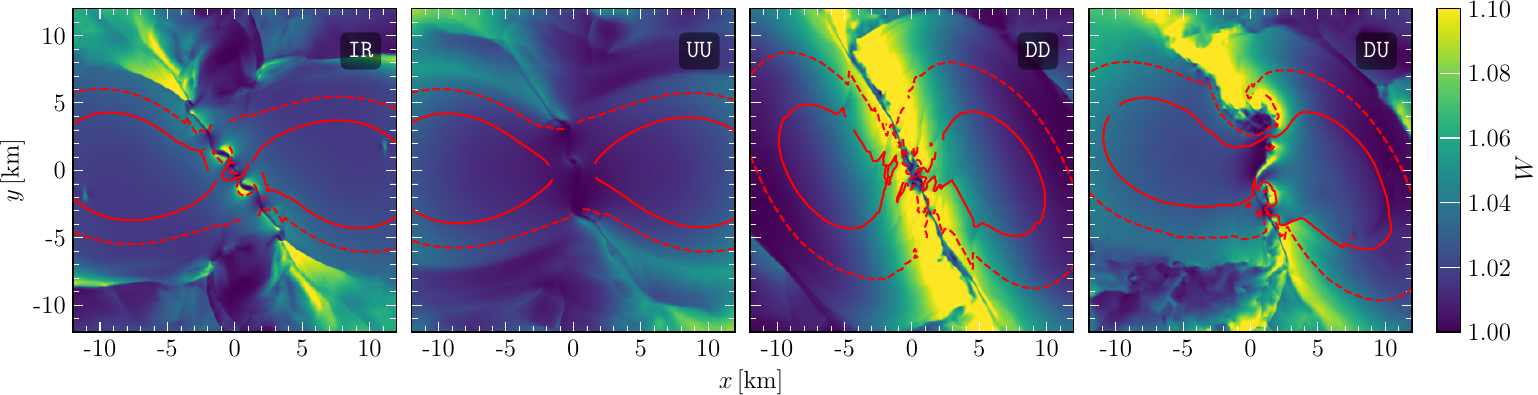}
  \caption{Distributions of the Lorentz factor $W$ for all cases
    considered (different columns) at time $t = t_{\rm mer}$. As in
    Fig.~\ref{fig:fig2}, the distributions are rotates so that the
    stellar centres are on a $y=0$ line. Dashed and solid lines
    correspond to the rest-mass density contours of $4\times 10^{14}$ and
    $6\times 10^{14}\,{\rm g~cm^{-3}}$, respectively. }
  \label{fig:fig3}
\end{figure*}

Hence, the first column in Fig.~\ref{fig:fig2} allows one to observe the
preliminary contacts between the tidally disrupted low-density shear
layers in the \texttt{DD} and \texttt{DU} cases. Note therefore that the
KHI can develop already early-on and in low-density matter; this result
modifies the expectation that shear layers prone to the KHI form
primarily in high-density regions, whereas the Rayleigh-Taylor
instability occurs in low-density matter~\citep{Palenzuela_2022PRD}. We
should also note that these early exponential growths by the KHI are not
observed in low-resolution simulations, indicating that, given
sufficiently high resolution, strong spin-induced dynamical tides can
trigger an earlier KHI phase. Specifically, in the \texttt{DU}
configuration, the tidal forces acting on the anti-aligned-spin star are
stronger than those on the aligned-spin star, causing it to plunge into
its companion and generate a shear layer earlier than in all other
binaries. Another aspect that can be appreciated from the first column of
Fig.~\ref{fig:fig2} is that the contact surface area can be quite
different and is systematically larger for the configurations involving
down-spin star(s). Indeed, at the moment the shear layer forms, only the
stellar tips begin to touch in the \texttt{IR} and \texttt{UU} cases. By
contrast, for the \texttt{DD} and \texttt{DU} configurations, which
experience stronger tidal deformations, the stellar bodies of two NSs
contact each other, resulting in a significantly larger stellar surface
area being involved in the collision.

Proceeding in time, the second and third columns of Fig.~\ref{fig:fig2}
gives a revealing representation of the magnetic-field distribution
around the moment of merger and when the KHI has been quenched. In this
way, it is possible to appreciate that the \texttt{DD} binary (third
row), which can benefit from a number of favourable factors, such as the
highest approaching velocity $v_a$ resulting from the smallest
angular-momentum loss (``hang-up'' effect) the smallest total angular
momentum (see the values of $v_a$ in Tab.~\ref{tab:ID} and the discussion
in Fig.~\ref{fig:fig3}), an earlier tidal shearing, a larger contact
surface area, and a larger discontinuity in the tangential velocities
(see arrows in Fig.~\ref{fig:fig2}) can lead to a very strong vorticity,
with eddies being produced in both the high- and low-density
regions. More importantly, the vorticity pervades all regions of the NSs,
penetrating down to the highest-density cores. This behaviour should be
contrasted with that reported for the traditional irrotational binary
\texttt{IR} (first row), where it is apparent that the vorticity at this
stage in the evolution is confined to the low-density regions of the two
stars~\citep[see also Fig.~2 of][for a snapshot at a later
  time]{Chabanov2022}. On the other extreme, the \texttt{UU} binary
(second row) has properties that are the opposite of those illustrated
for the binary \texttt{DD} and thus develops a much weaker vorticity as a
result of milder kinetic conditions at the time of the merger~\citep[see
  also][]{Neuweiler2025b}, with significant velocity gradients appearing
only when the high-density regions collide. Indeed, the \texttt{UU}
binary exhibits the longest inspiral and the strongest GW emission, which
results in a deficit of both kinetic and internal energies, producing the
few and weak eddies and ultimately forming a colder remnant~\citep[see
  also][]{Karakas2024}. Given the very different conditions at merger, it
is not surprising that the \texttt{DD}/\texttt{UU} binaries lead to the
largest/smallest overall growth in the EM energy right after KHI phase.

A special discussion should be reserved for the \texttt{DU} binary
(fourth row), where the turbulence is initially developed and is
restricted only to the NS with anti-aligned spin~\citep[a similar
  behaviour is shown in Fig.~1 of][although a detailed discussion is not
  presented]{Aguilera-Miret2025}. To understand the origin of this
behaviour, it is useful to recall that the NS with anti-aligned spin is
severely tidally deformed during the late inspiral and that the material
that is stripped forms a strong shear layer surrounding the NS with
aligned-spin. (see also Fig.~\ref{fig:fig3}). In turn, this leads to a
very asymmetrical flow and vorticity distribution which leads to the
behaviour described in the last row of Fig.~\ref{fig:fig2}. As a result,
although the \texttt{DU} binary has the second-largest approaching
velocity and generates a significant fraction of vortices during the
early phase of the tidal shearing, the large asymmetry in the flow
prevents a direct collision between the two NSs and an efficient KHI
development. The asymmetry also disrupts the shear layer and very
effectively suppresses the development of eddies. This behaviour also
explains why the growth of the EM energy in the \texttt{IR} binary
becomes larger than that of the \texttt{DU} binary at later times (see
top panel of Fig.~\ref{fig:fig1}).

The variations in behaviours among all cases discussed, such as the
collision strength, the tidal dynamics, and the contact area, clearly
reveal that, when sufficiently high resolutions are employed, different
spin configurations create intrinsically different dynamics and
substantially contrasting environments for the development of KHI-driven
eddies. The fact that the early post-merger phase is sensitive -- in
addition to spin -- also to distinct magnetic-field configurations or
strengths~\citep{Chabanov2022, Gutierrez2025}, is not necessarily in
contrast with the suggestion that a quasi-universal postmerger
magnetic-field state will be reached at sufficiently high
resolutions~\citep[see, \eg][]{Aguilera-Miret2021}. It is indeed possible
that much of the initial conditions -- although not all of them -- will
be washed out if the evolution takes place for sufficiently long
timescales. What is relevant to remark, however, is that we clearly show
that initial spins -- more than the initial magnetic-field -- can largely
alter the intrinsic dynamics of the system and drive different
strength/population of KHI eddies, which are not affected by simply
changing the magnetic-field strength and topology. Therefore, our results
could form the postmerger remnants with significantly different saturated
states as a result of the different developments of the KHI. In turn, the
different magnetic-field amplifications can lead EM emission in this
time-frame that will be obviously different.

As anticipated above, the initial spin in the two stars regulates the
angular momentum of the system at merger and thus determines the amount
of kinetic energy with which the two stars collide. In turn, the
different speeds at which the star shear -- and the strength of the
shocks resulting from the collision -- will play a fundamental role in
shaping the vorticity and thus the efficiency of the KHI in amplifying
the magnetic field in the four scenarios considered. In this sense, the
magnetic-field amplification shown in Fig.~\ref{fig:fig1} reflects the
combined effects of collision properties, the extent of vorticities
generated, and the survival time of the shear layer.

To illustrate the role played by shocks at merger, Fig.~\ref{fig:fig3}
shows the Lorentz factor in the Eulerian frame $W$ at time $t=t_{\rm
  mer}$, together with isocontours of the rest-mass density to represent
distribution of rest-mass as the two NSs collide. Taking the distribution
of the Lorentz factor as a proxy for the strength of the shocks at
merger, Fig.~\ref{fig:fig3} explains why the \texttt{UU} binary
experiences the longest KHI but also has the lowest magnetic-field
magnification. This is because, at merger, this binary generates the
weakest vorticity but also experiences the weakest shocks after collision
(second panel from the left). By contrast, the NSs in the \texttt{DD}
binary approach with a large kinetic energy, produce very strong shocks
(second panel from the right) and these can expose the dense regions of
the stars -- that are obviously endowed with the strongest magnetic
fields -- to the vorticity of the KHI.

From Fig.~\ref{fig:fig3} it is also possible to appreciate the special
nature of the \texttt{DU} binary (right panel), which experiences an
asymmetric but strong shock at collision. This behaviour, together with
the strong but limited vorticity discussed in Fig.~\ref{fig:fig2}, leads
to a large decay in the growth rate (see Fig.~\ref{fig:fig1}), giving
this binary -- together with \texttt{DU} -- the shortest lifetime among
those with nonzero spin. Finally, also noticeable in Fig.~\ref{fig:fig3}
is the different extents of the surface areas at contact. In particular,
note how all stars with an anti-aligned spin (\texttt{DD} and \texttt{DU})
exhibit substantially larger contact surface areas. By contrast, the
binary with aligned spins (\texttt{UU}) has the smallest effective
surface at the time of the collision.

Finally, Fig.~\ref{fig:fig3} also helps in understanding the differences
in the duration of the KHI, which is the result of the interplay between
the generated shocks, the survival time of the shearing region, and the
strength and extent of the vorticity eddies. Table~\ref{tab:ID} reports
the KHI duration time, which is defined as $t_{\rm KHI} := t_{\rm sh,
  1/2} - t_{\rm sh}$, and it is then possible to realise that the
\texttt{UU} binary, which we have already commented leads to the weakest
amplification, also experiences the longest-lasting KHI ($t_{\rm KHI} =
1.35~\rm{ms}$). On the contrary, the \texttt{DD} binary, which leads to
the strongest amplification has a rather short-lived KHI ($t_{\rm KHI} =
1.01~\rm{ms}$). Interestingly, while the \texttt{IR} binary exhibits the
second strongest collision strength (excluding the special \texttt{DU}
case), it possesses the shortest-lived KHI. This is mostly because it
lacks the the extended period of early shearing driven by the strong
dynamical tidal effects present in anti-aligned spin configurations.

\begin{figure}[t!]
  \includegraphics[width=0.48\textwidth]{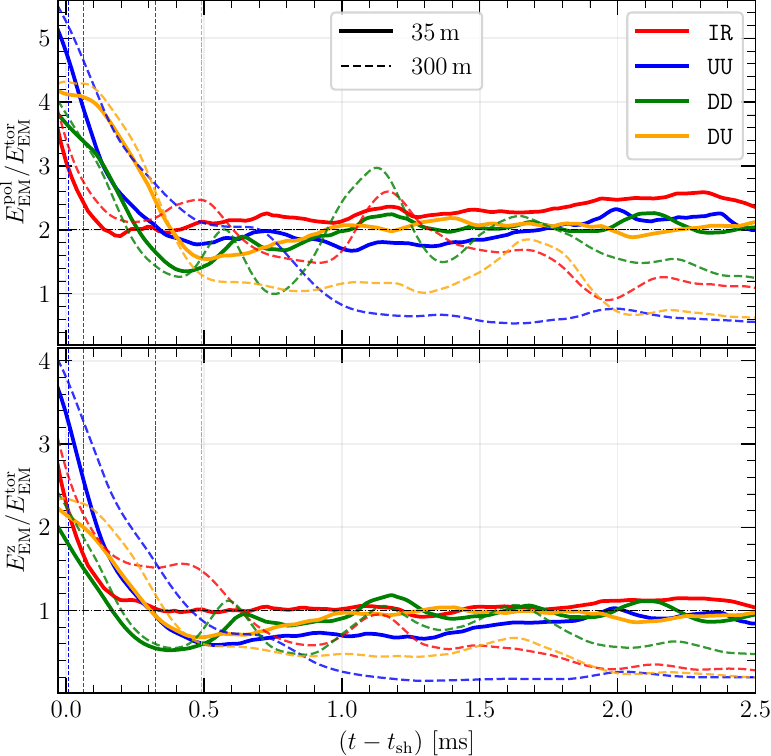}
  \caption{\textit{Top panel:} evolution of the ratio of the EM energy in
    the poloidal and toroidal components (solid lines) for all binaries
    considered; note that at late times the ratio converges to a constant
    value of $\simeq 2$. \textit{Bottom panel:} as in the top but
    relative to the ratio between the $z$- and the toroidal components;
    note that at late times the ratio converges to a constant value of
    $\simeq 1$. Shown with dashed lines are the same quantities when
    employing a coarser resolution of $300\,{\rm m}$ and yielding to
    significantly different values.}
  \label{fig:fig4}
\end{figure}
%

\subsection{On the Topological Equipartition}

The study of vorticity in BNS mergers has a long history [see, \eg Fig.~5
  of \citet{Rasio99__}, Fig.~3 of~\citet{Price06}, or Fig.~16
  of~\citet{Baiotti08}] and has a direct impact on the evolution of the
topological components of the magnetic field, \ie the toroidal and
poloidal one. Given the purely (or predominantly) poloidal nature of the
initial magnetic field and the predominantly toroidal nature of the flow
before merger, the induction equation in ideal MHD trivially predicts the
growth of the toroidal component. Indeed, this is also what is recorded
in Fig.~\ref{fig:fig1}, whose fourth panel from the top shows how the
toroidal component has the largest growth rate initially right after the
formation of shear.

Yet, also the poloidal component of the magnetic energy grows of three
orders of magnitude from the initial values and this is less obvious to
account for. \cite{Palenzuela_2022PRD} and \citet{Kiuchi2023} briefly
note that KHI-induced small-scale turbulence is isotropic, amplifying
magnetic fields in random orientations and thereby producing similar
growth rates for both poloidal and toroidal components. While this is
correct as a first approximation, it does not explain why the growth
rates of the toroidal and poloidal components are offset in time, with
the latter (former) prevailing at later (earlier) times. As we discuss
below, this behaviour can be understood because the KHI-induced eddies
are initially anisotropic -- \ie confined to specific regions of the star
where the shearing takes place -- and only later spread to the stellar
interior, with the initial spins playing an important role in determining
when the transition to isotropy takes place. Furthermore, after the
transition has taken place, there is no topological equipartition between
the toroidal and poloidal components, \ie $E_{_{\rm EM}}^{\rm{pol}} /
E_{_{\rm EM}}^{\rm tor} \neq 1$, although there is an equipartition in
the vertical and horizontal directions for all binaries considered,
despite their dynamics is considerably different.

To illustrate this finding, we show in Fig.~\ref{fig:fig4} the evolution
of the ratio between the EM energy in the poloidal and toroidal
components $E_{_{\rm EM}}^{\rm{pol}} / E_{_{\rm EM}}^{\rm tor}$ (top
panel), as well as of the ratio between the vertical [\ie in the
  $z$-direction when adopting a decomposition in cylindrical coordinates
  $(\varpi, \phi, z)$] and the horizontal (\ie toroidal) components
$E_{_{\rm EM}}^{z} / E_{_{\rm EM}}^{\rm tor}$ (bottom panel). We note
that the vertical component is produced by the generation of vertical
vortices at the time of contact of the high-density regions (see also
Appendix~\ref{sec:app_A}) and that this component is generated at
different times for different binaries. In particular, this poloidal
component will grow after the toroidal one and as a result of shearing
away from the equatorial plane (see also bottom panel of
Fig.~\ref{fig:fig1}). Lines of different colour refer to the different
binaries, and solid lines are used to report the results of the
high-resolution simulations ($35\,{\rm m}$), while the dashed lines refer
to the low-resolution ones ($300\,{\rm m}$); here too, the vertical
dashed lines mark $t_{\rm mer}$.

Collecting the information in the two panels of Fig.~\ref{fig:fig4}, it
then becomes clear that when the KHI has been quenched, for all four
binaries $E_{_{\rm EM}}^{\rm{pol}} \sim 2 E_{_{\rm EM}}^{\rm tor}$, but
also that
\begin{equation}
\label{eq:hierarchy}
  E^{\varpi}_{_{\rm EM}} \approx E^z_{_{\rm EM}} \approx E^{\rm
    tor}_{_{\rm EM}} \approx E^{\rm pol}_{_{\rm EM}}/2 \,,
\end{equation}
where we recall that $E^{\rm pol}_{_{\rm EM}} := E^{\varpi}_{_{\rm EM}} +
E^z_{_{\rm EM}}$. In other words, after the KHI has ended, equipartition
in all magnetic-fields components is reached and the poloidal component
will always be twice as large as the toroidal component. Importantly,
this result, i.e., the equivalence between $E_{_{\rm EM}}^{z}$ and
$E_{_{\rm EM}}^{\rm tor}$, can be attained only when sufficiently large
resolutions are used, because low-resolution simulations do not show a
common behaviour (dashed lines in Fig.~\ref{fig:fig4}).

\section{Conclusions}

Assessing if and how the comparatively low magnetic fields of $\sim
10^{8}-10^{10}\,{\rm G}$ characterising two NSs before their merger are
amplified to magnetar-strengths of $\sim 10^{15}-10^{16}\,{\rm G}$ only a
couple of milliseconds after merger represents one of the most important
aspects of the dynamics of BNS systems and remains only poorly
understood. The enormous computational costs associated with the direct
numerical simulations that can probe reliably this process have so far
prevented an exhaustive investigation of the impact that the initial
spins have on the merger dynamics and on the development of the
Kelvin-Helmholtz instability (KHI).

Leveraging on general-relativistic magnetohydrodynamic simulations at
very high-resolution of $35\,{\rm m}$ and on a novel strategy exploiting
the combined approach of two numerical-relativity codes, we have here
reported the first systematic study of how rapid rotation impact the
KHI-amplification process. The strategy consists of employing a
numerical-relativity GRMHD code for the solution at comparatively smaller
resolutions of the full set of Einstein equations in the (long) inspiral
phase and to ``\HO'' the data over to different code that can solve them
in the xCFC approximation at much higher resolutions. While this approach
is still computationally challenging, it yields a substantial
computational saving that makes a systematic investigation as this one
effectively possible.

To explore sparsely but broadly the space of parameters, we have
considered four different spinning configurations that include, besides a
standard irrotational binary, also binaries with rapid spin
configurations that are either aligned, anti-aligned,
aligned/anti-aligned (mixed) with respect to the orbital angular
momentum. In this way, and contrary to previous expectations deduced at
much lower resolutions that the spin contributions would not affect the
KHI, we have found that the KHI-induced amplification of the magnetic
field, which represents only a portion of the window in time covered by
the simulations, is strongly influenced by the spin of merging stars.

More specifically, the binary with aligned spins (\texttt{UU}) has, at
the end of the simulations, the lowest EM energy considered
among all cases considered. By contrast, the binary with anti-aligned
spins (\texttt{DD}) has an EM energy that is $\sim 5$ times larger than
for the \texttt{UU} binary and is the largest of the ones
found. Interestingly, even though the evolution is rather different, the
mixed-spin binary (\texttt{DU}) has a final EM energy that is comparable
with that of the irrotational binary (\texttt{IR}).

These differences are the consequence of a number of intrinsic variations
in the system dynamics, which include: the tidal deformation before
merger, the jump in the tangential velocity -- and hence the vorticity --
at contact, the shock strength at the collision, and the extent of the
surface area involved in the KHI. At the cost of oversimplifying, the
role of spin in the development of the KHI can be summarised as follows:
in order to obtain a strong magnetic-field amplification via the KHI it
is necessary to have both strong shocks (that can expose strong magnetic
fields) and strong vorticity (that can mix them efficiently). The
\texttt{DD} binary experiences both strong shocks and strong vorticity,
and is thus the most efficient amplifier. The \texttt{DU} binary
experiences strong shocks but has an asymmetric vorticity, which reduces
the efficiency in the amplification. The \texttt{UU} binary experiences
neither strong shocks nor strong vorticity, making it the least efficient
magnetic-field amplifier among the configurations considered.

Our simulations, and the high-resolutions at which they are performed,
also reveal that despite the fact that different spin configurations
yield significantly different initial growth rates of the
poloidal/toroidal components, all binaries converge to a precise
topological partition as a result of the formation of vertical vortices,
advection and cascade of the EM energy in small-scale eddies. More
specifically, after the KHI has ended, the poloidal component of the EM
energy is always twice as large as the toroidal one, with true
equipartition being reached only in the three components of the magnetic
field.

It is possible that the dynamics described here -- and in particular the
topological equipartition -- is in large part ``quasi-universal'', that
is, largely independent of the system properties, such as the EOS, the
mass ratio, the initial magnetic-field and the binary eccentricity. This
is because, for sufficiently high rotation, the spin-induced corrections
discussed here in the development of the KHI will remain the most
important in the budget of possible effects influencing the KHI. New
simulations that are longer, explore different mass ratios and EOSs, and
include neutrino transport are necessary to confirm this conjecture.

\label{sec:summary}

\begin{acknowledgments}
  It is a pleasure to thank M. Cassing, M. Chabanov, E. R. Most, and
  K. Topolski for useful discussions. This research is supported by the
  ERC Advanced Grant ``JETSET: Launching, propagation and emission of
  relativistic jets from binary mergers and across mass scales'' (grant
  No. 884631), by the Deutsche Forschungsgemeinschaft (DFG, German
  Research Foundation) through the CRC-TR 211 ``Strong-interaction matter
  under extreme conditions'' -- project number 315477589 -- TRR
  211. L.R. is grateful to the Theory Division at CERN, where part of
  this research was carried out and to the Walter Greiner Gesellschaft
  zur F\"orderung der physikalischen Grundlagenforschung e.V. through the
  Carl W. Fueck Laureatus Chair.
\end{acknowledgments}

\clearpage\newpage

\appendix

\begin{figure*}[t!]
  \includegraphics[width=0.98\textwidth]{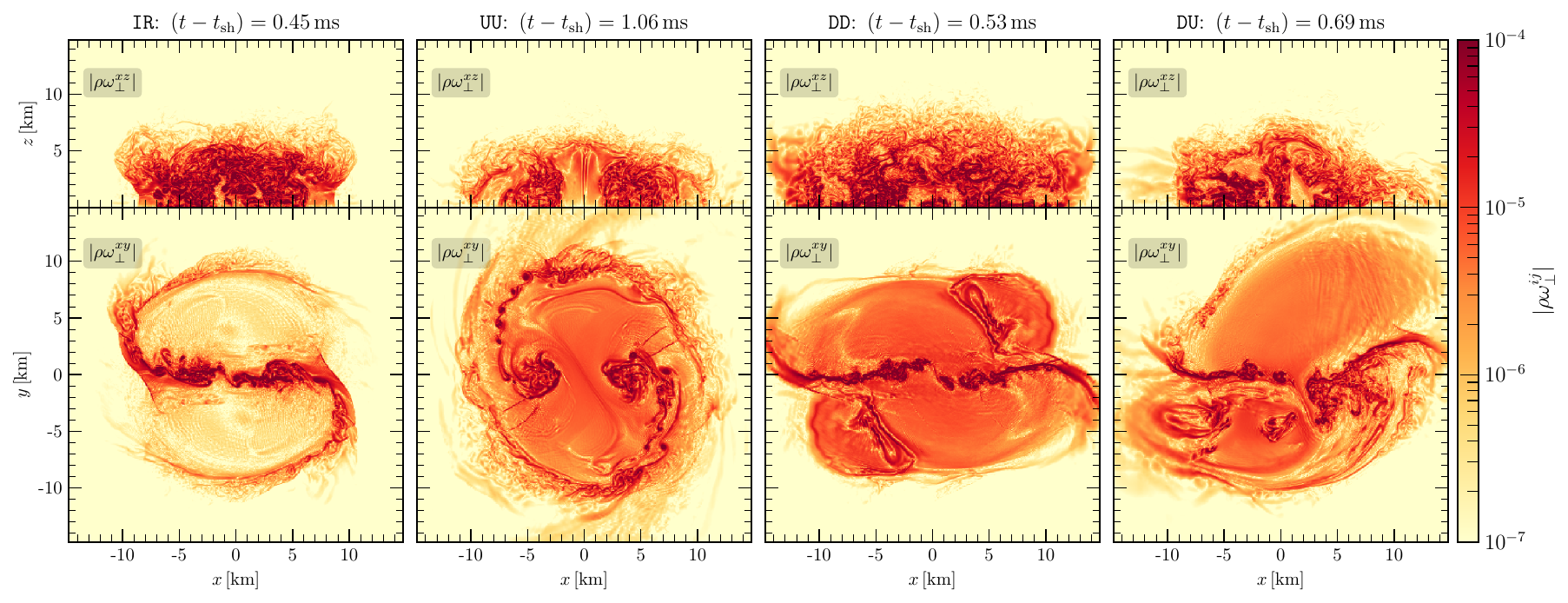}
  \caption{Absolute values of density-weighted kinematic vorticity
    $\left|\rho \omega_{\perp}^{ij}\right|$ for all cases at the time
    when the growth rate of poloidal EM energy reach their peaks. The top
    and bottom rows show the distributions of $\left|\rho
    \omega_{\perp}^{xz}\right|$ and $\left|\rho
    \omega_{\perp}^{xy}\right|$ on the $(x,z)$ and $(x,y)$ planes,
    respectively. The different columns refer to the four binaries
    considered and at four representative times. The distributions are
    rotated so as to highlight the different distributions of vorticity.}
      \label{fig:fig5}
\end{figure*}

\section{Vertical and Horizontal Vorticity}
\label{sec:app_A}

\begin{figure}[t!]
  \centering
  \includegraphics[width=0.40\textwidth]{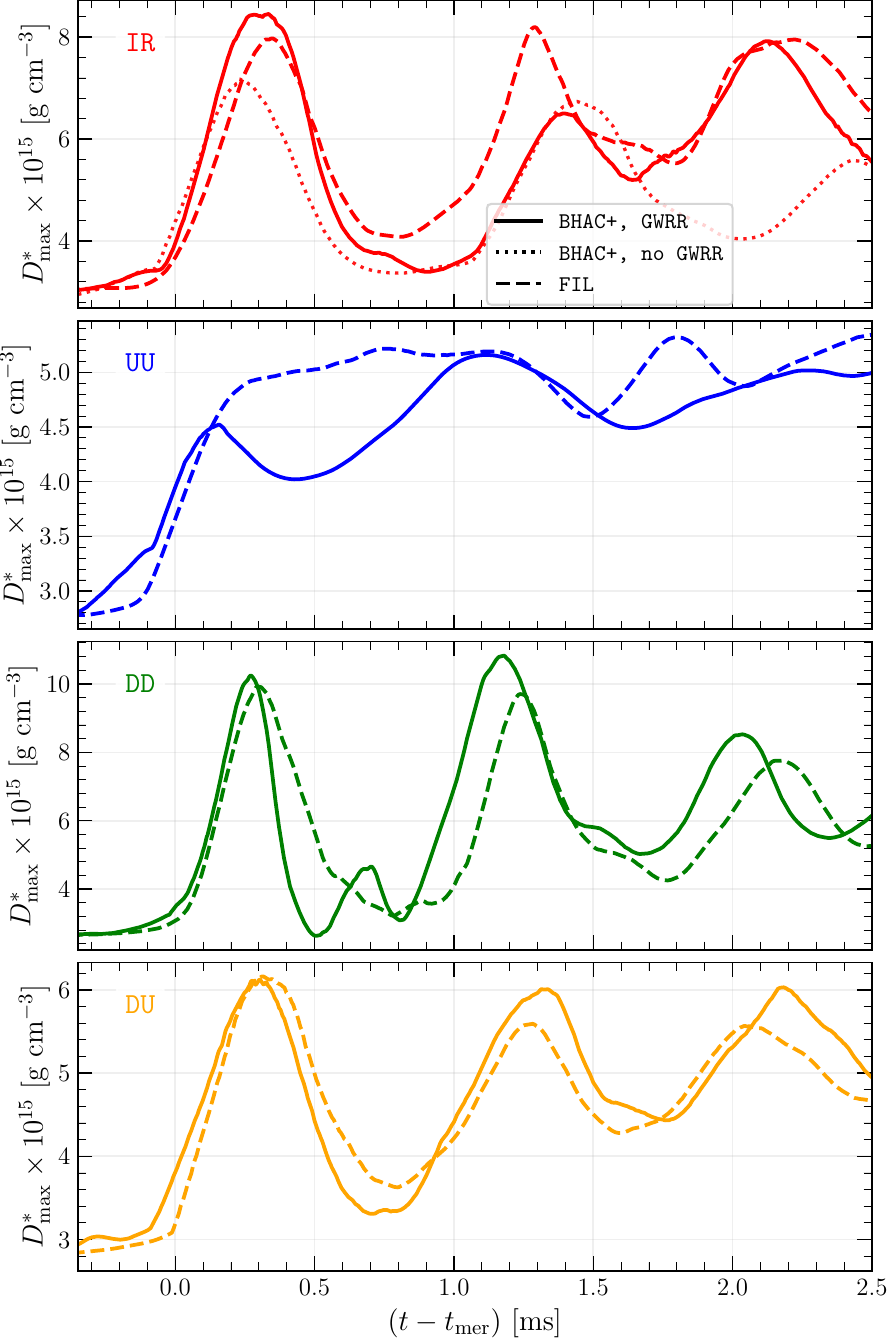}
  \caption{Time evolution of the maximum of the conserved rest-mass
    density from simulations using \FIL (dashed lines) and \BHAC, either
    without GWRR terms (dotted lines) or with them included. Note
    the importance of the latter in reproducing the results of overall dynamics comparing 
    to the full numerical-relativity simulations. In all cases, the finest spatial
    resolution is of $300\, \rm m$ resolution. }
  \label{fig:fig6}
\end{figure}

In the main text we have illustrated how, at late times, the evolution of
the ratio of the EM energy in the poloidal and toroidal components
converges to a constant value $E^{\rm tor}_{_{\rm EM}} \approx E^{\rm
  pol}_{_{\rm EM}}/2$ (see Fig.~\ref{fig:fig4}). In what follows we
provide a number of elements to explain this result. To this scope, we
use Fig.~\ref{fig:fig5}, to show for the four binaries considered the
spatial distribution of the absolute value of the (rest-mass)
density-weighted kinematic vorticity. We recall that the kinematic
vorticity is defined as
\begin{equation}
\omega_{\perp}^{x y}:=h^x_{~\mu} h^y_{~\nu} \omega^{\mu \nu}\,,
\end{equation}
where $\omega_{\mu \nu}$ and $h_{\mu\nu} := g_{\mu\nu} + u_{\mu}u_{\nu}$
are the vorticity and projector tensor orthogonal to the fluid
four-velocity $u^{\mu}$, respectively~\citep{Rezzolla_book:2013}.
Figure~\ref{fig:fig5} reports the ``vertical vorticity'', \ie the
weighted vorticity $\left|\rho \omega_{\perp}^{x z}\right|$ in the polar
plane (top row) and the ``horizontal vorticity, \ie $\left|\rho
\omega_{\perp}^{x y}\right|$ in the equatorial planes (bottom row).
Since the vorticity tensor has all components being nonzero, the split in
vertical and horizontal vorticity is, to some extent, arbitrary. However,
such simple split helps to understand the evolution of the growth rates
in the toroidal and poloidal components. Indeed, the strength of
$\left|\rho \omega_{\perp}^{xy}\right|$ and $\left|\rho
\omega_{\perp}^{xz}\right|$ can be taken as proxies for the strength of
$\gamma^{\rm tol}$ and $\gamma^{\rm pol}$, respectively.

The selected snapshots correspond to the times when the growth rate of
the poloidal EM energy reaches its maximum, and in all cases these times
occur for $t_{\rm sh, 1/2} > t > t_{\rm mer}$. While at these times the
vertical vorticity is comparable to the horizontal vorticity, earlier on
in the shearing (not shown in Fig.~\ref{fig:fig5}), the vorticity is
mostly horizontal as the high-density regions of the two stars have not
yet come into contact and the shear develops mostly with the low-density
regions of the two stars. Indeed, a vertical vorticity is hardly present
in the early-shear stages of the \texttt{DD} and \texttt{IR} binaries, so
that for these binaries the growth rate of the toroidal component is
significantly higher than that of the poloidal one (see also
Fig.~\ref{fig:fig1}). At the times depicted in Fig.~\ref{fig:fig5}, and
also subsequently, the vertical vorticity will develop further as the
collision of the high-density regions squeezes and moves upward the
matter compressed by the colliding stellar cores. Hence, the poloidal
component of the EM energy will experience an exponential growth
immediately after $t = t_{\rm mer}$.

This dichotomy in the development of the turbulence, \ie mostly
horizontal at early times but then also vertical at later times
(especially for the \texttt{DD} and \texttt{IR} binaries), and the
corresponding anisotropy in the turbulence distribution, is washed out at
later times, when the turbulence becomes increasingly isotropic. This is
possible because the magnetic feedback on the matter dynamics remains
negligible, since the amplifying field is still relatively weak ($\sim
10^{11}$--$10^{13}\,{\rm G}$). The reaching of an isotropic turbulence
such that the strength of magnetic field is equally distributed in all
the components (topological equipartition), provides then the natural
explanation for the energy hierarchy expressed by
Eq.~\eqref{eq:hierarchy} and represented in the top panel of
Fig.~\ref{fig:fig4}.

\section{Comparing xCFC-GWRR with full general relativity}
\label{sec:app_B}

The use of an approximate gravity solver -- based on xCFC approximation
and combined with GWRR terms -- is what allowed us to perform
calculations at high resolution and affordable computational
costs. Clearly, it is essential to establish what is the error that such
approximation introduces and this can be easily done when comparing with
a full general-relativistic simulations performed at a low-resolution of
$300\,\rm m$. Figure~\ref{fig:fig6} shows the evolution of the maximum of
the conserved rest-mass density~\citep{Rezzolla_book:2013}, $D^{*} :=
\rho W \psi^6$ for $2.5\,\rm ms$ after merger [here $\psi$ is the
  conformal factor in the xCFC formulation~\citep{Jiang2025}], comparing
results from \BHAC with GWRR (solid lines), and \FIL (dashed lines) for
all the binaries simulated. To show the importance of the GWRR
corrections, we show the evolution without such terms (dotted line) in
the case of the \texttt{IR} binary (top panel). Note the considerable
differences that emerge for $t-t_{\rm mer} > 1.5~\rm{ms}$.

Despite the fundamental differences between the constrained-evolution
scheme with CFC and the constraint-damping Z4 scheme, the two codes show
an agreement that is overall very good. Only a slight phase difference of
approximately $0.1$-$0.2\,\rm ms$ develops over time, and in \BHAC the
two stellar cores collide slightly earlier (by $\sim 0.05\,\rm ms$),
which we attribute to the metric initialisation under the CFC
approximation. Across the four test cases, the average relative
differences range from $\sim 3\%$ to $12\%$, while the maximum deviations
lie between approximately $10\%$ and $25\%$. We should note that this is
a very critical test for the xCFC-GWRR scheme, which was shown to be very
accurate in the post-merger phase, when the remnant and the spacetime
have reached an approximate axisymmetry~\citep{Jiang2025}. The dominant
source of discrepancy arises from the GWRR treatment, which corrects the
GW energy and angular-momentum loss in the xCFC scheme up to $2.5$~PN
order but only in the lapse function, leaving the shift vector and the
non-diagonal components of the spatial metric
unchanged~\citep{Jiang2025}.  It should also be emphasised that
differences of this order (or even larger) have been reported between the
post-merger evolution computed when employing different but fully general
relativistic codes~\citep{Espino2023}.

One may be concerned that the differences in kinetic and thermal energies
introduced by the approximate gravity treatment could influence the onset
and development of turbulence. However, because the fraction of kinetic
energy converted into EM energy is only $\sim 1\%$, the
observed relative differences of $\lesssim 10\%$ remain largely
subdominant over the resolution-induced errors that govern the dynamics
of the KHI.

As a final comment, we note that to minimise the computational costs, we
follow the strategy of~\cite{Jiang2025} and optimise the ratio $\chi :=
\Delta t_{\rm met} / \Delta t_{\rm MHD}$, where $\Delta t_{\rm met}$ is
the interval between two metric updates and $\Delta t_{\rm MHD}$ is the
MHD timestep.  Since the relative difference between BHAC and FIL
solutions reduces to $\lesssim 0.5\%$ for $\Delta t_{\rm met} \leq 5
\times 10^{-2} \, M_{\odot}$, we adopt $\Delta t_{\rm met} = 5 \times
10^{-2} \, M_{\odot}$ as the upper bound. This reduces the computational
costs considerably and provides a substantial benefit in high-resolution
runs where $\Delta t_{\rm MHD}$ becomes very small.

\bibliography{aeireferences}

\begin{thebibliography}{}
\expandafter\ifx\csname natexlab\endcsname\relax\def\natexlab#1{#1}\fi
\providecommand{\url}[1]{\href{#1}{#1}}
\providecommand{\dodoi}[1]{doi:~\href{http://doi.org/#1}{\nolinkurl{#1}}}
\providecommand{\doeprint}[1]{\href{http://ascl.net/#1}{\nolinkurl{http://ascl.net/#1}}}
\providecommand{\doarXiv}[1]{\href{https://arxiv.org/abs/#1}{\nolinkurl{https://arxiv.org/abs/#1}}}

\bibitem[{{Abbott} {et~al.}(2017){Abbott}, {Abbott}, {Abbott}, {Acernese},
  {Ackley}, {Adams}, {Adams}, {Addesso}, {Adhikari}, {Adya}, \&
  et~al.}]{Abbott2017_etal}
{Abbott}, B.~P., {Abbott}, R., {Abbott}, T.~D., {et~al.} 2017, Phys. Rev.
  Lett., 119, 161101, \dodoi{10.1103/PhysRevLett.119.161101}

\bibitem[{Acker {et~al.}(2016)Acker, B.~de R.~Borges, \& Costa}]{Acker2016}
Acker, F., B.~de R.~Borges, R., \& Costa, B. 2016, J. Comput. Phys., 313, 726,
  \dodoi{10.1016/j.jcp.2016.01.038}

\bibitem[{Aguilera-Miret {et~al.}(2025)Aguilera-Miret, Christian, Rosswog, \&
  Palenzuela}]{Aguilera-Miret2025}
Aguilera-Miret, R., Christian, J.-E., Rosswog, S., \& Palenzuela, C. 2025, Mon.
  Not. Roy. Astron. Soc., 3067, 3077, \dodoi{10.1093/mnras/staf1291}

\bibitem[{{Aguilera-Miret} {et~al.}(2024){Aguilera-Miret}, {Palenzuela},
  {Carrasco}, {Rosswog}, \& {Vigan{\`o}}}]{Aguilera-Miret2024}
{Aguilera-Miret}, R., {Palenzuela}, C., {Carrasco}, F., {Rosswog}, S., \&
  {Vigan{\`o}}, D. 2024, Phys. Rev. D, 110, 083014,
  \dodoi{10.1103/PhysRevD.110.083014}

\bibitem[{Aguilera-Miret {et~al.}(2022)Aguilera-Miret, Vigan\`o, \&
  Palenzuela}]{Aguilera-Miret2021}
Aguilera-Miret, R., Vigan\`o, D., \& Palenzuela, C. 2022, Astrophys. J. Lett.,
  926, L31, \dodoi{10.3847/2041-8213/ac50a7}

\bibitem[{{Alic} {et~al.}(2012){Alic}, {Bona-Casas}, {Bona}, {Rezzolla}, \&
  {Palenzuela}}]{Alic:2011a}
{Alic}, D., {Bona-Casas}, C., {Bona}, C., {Rezzolla}, L., \& {Palenzuela}, C.
  2012, Phys. Rev. D, 85, 064040, \dodoi{10.1103/PhysRevD.85.064040}

\bibitem[{{Baiotti} {et~al.}(2008){Baiotti}, {Giacomazzo}, \&
  {Rezzolla}}]{Baiotti08}
{Baiotti}, L., {Giacomazzo}, B., \& {Rezzolla}, L. 2008, Phys. Rev. D, 78,
  084033, \dodoi{10.1103/PhysRevD.78.084033}

\bibitem[{Baiotti \& Rezzolla(2017)}]{Baiotti2016}
Baiotti, L., \& Rezzolla, L. 2017, Rept. Prog. Phys., 80, 096901,
  \dodoi{10.1088/1361-6633/aa67bb}

\bibitem[{Bamber {et~al.}(2024)Bamber, Tsokaros, Ruiz, \& Shapiro}]{Bamber2024}
Bamber, J., Tsokaros, A., Ruiz, M., \& Shapiro, S.~L. 2024, Phys. Rev. D, 110,
  024046, \dodoi{10.1103/PhysRevD.110.024046}

\bibitem[{{Bauswein} \& {Stergioulas}(2015)}]{Bauswein2015}
{Bauswein}, A., \& {Stergioulas}, N. 2015, Phys. Rev. D, 91, 124056,
  \dodoi{10.1103/PhysRevD.91.124056}

\bibitem[{Bernuzzi \& Hilditch(2010)}]{Bernuzzi:2009ex}
Bernuzzi, S., \& Hilditch, D. 2010, Phys. Rev. D, 81, 084003,
  \dodoi{10.1103/PhysRevD.81.084003}

\bibitem[{{Blanchet}(2014)}]{Blanchet2014}
{Blanchet}, L. 2014, Living Reviews in Relativity, 17, 2,
  \dodoi{10.12942/lrr-2014-2}

\bibitem[{{Bonazzola} \& {Gourgoulhon}(1996)}]{Bonazzola1996}
{Bonazzola}, S., \& {Gourgoulhon}, E. 1996, Astron. and Astrophys., 312, 675

\bibitem[{{Bovard} {et~al.}(2017){Bovard}, {Martin}, {Guercilena}, {Arcones},
  {Rezzolla}, \& {Korobkin}}]{Bovard2017}
{Bovard}, L., {Martin}, D., {Guercilena}, F., {et~al.} 2017, Phys. Rev. D, 96,
  124005.
\newblock \doarXiv{1709.09630}

\bibitem[{{Bucciantini} \& {Del Zanna}(2011)}]{Bucciantini2011}
{Bucciantini}, N., \& {Del Zanna}, L. 2011, Astron. Astrophys., 528, A101,
  \dodoi{10.1051/0004-6361/201015945}

\bibitem[{Celora {et~al.}(2025)Celora, Palenzuela, Vigan{\`o}, \&
  Aguilera-Miret}]{Celora2025}
Celora, T., Palenzuela, C., Vigan{\`o}, D., \& Aguilera-Miret, R. 2025.
\newblock \doarXiv{2505.01208}

\bibitem[{{Chabanov} {et~al.}(2023){Chabanov}, {Tootle}, {Most}, \&
  {Rezzolla}}]{Chabanov2022}
{Chabanov}, M., {Tootle}, S.~D., {Most}, E.~R., \& {Rezzolla}, L. 2023,
  Astrophys. J. Lett., 945, L14, \dodoi{10.3847/2041-8213/acbbc5}

\bibitem[{{Chen} {et~al.}(2025){Chen}, {Zhang}, {Wang}, {Tan}, {Xiong}, {Yang},
  {Yin}, {Zhang}, \& {Zhang}}]{Chen2025b}
{Chen}, R.-C., {Zhang}, B.-B., {Wang}, C.-W., {et~al.} 2025, Nature Astronomy,
  \dodoi{10.1038/s41550-025-02649-w}

\bibitem[{Cheong {et~al.}(2025)Cheong, Foucart, Ng, Offermans, Duez, Muhammed,
  \& Chawhan}]{Cheong2024c}
Cheong, P. C.-K., Foucart, F., Ng, H. H.-Y., {et~al.} 2025, Phys. Rev. D, 111,
  043036, \dodoi{10.1103/PhysRevD.111.043036}

\bibitem[{Cheong {et~al.}(2021)Cheong, Lam, Ng, \& Li}]{Cheong2021}
Cheong, P. C.-K., Lam, A. T.-L., Ng, H. H.-Y., \& Li, T. G.~F. 2021, Monthly
  Notices of the Royal Astronomical Society, 508, 2279,
  \dodoi{10.1093/mnras/stab2606}

\bibitem[{{Chi-Kit Cheong} {et~al.}(2024){Chi-Kit Cheong}, {Tsokaros}, {Ruiz},
  {Venturi}, {Chan}, {Yip}, \& {Uryu}}]{Cheong2024d}
{Chi-Kit Cheong}, P., {Tsokaros}, A., {Ruiz}, M., {et~al.} 2024, arXiv
  e-prints, arXiv:2409.10508, \dodoi{10.48550/arXiv.2409.10508}

\bibitem[{Ciolfi(2020)}]{Ciolfi2020}
Ciolfi, R. 2020, Monthly Notices of the Royal Astronomical Society: Letters,
  495, L66, \dodoi{10.1093/mnrasl/slaa062}

\bibitem[{{Combi} \& {Siegel}(2023)}]{Combi2022}
{Combi}, L., \& {Siegel}, D.~M. 2023, Astrophys. J., 944, 28,
  \dodoi{10.3847/1538-4357/acac29}

\bibitem[{Combi \& Siegel(2023)}]{Combi2023}
Combi, L., \& Siegel, D.~M. 2023, Phys. Rev. Lett., 131, 231402,
  \dodoi{10.1103/PhysRevLett.131.231402}

\bibitem[{Cook {et~al.}(2025)Cook, Guti{\'e}rrez, Bernuzzi, Radice, Daszuta,
  Fields, Hammond, Bandyopadhyay, \& Jacobi}]{Cook2025}
Cook, W., Guti{\'e}rrez, E.~M., Bernuzzi, S., {et~al.} 2025.
\newblock \doarXiv{2508.19342}

\bibitem[{{Cordero-Carri{\'o}n} {et~al.}(2009){Cordero-Carri{\'o}n},
  {Cerd{\'a}-Dur{\'a}n}, {Dimmelmeier}, {Jaramillo}, {Novak}, \&
  {Gourgoulhon}}]{Cordero2009}
{Cordero-Carri{\'o}n}, I., {Cerd{\'a}-Dur{\'a}n}, P., {Dimmelmeier}, H.,
  {et~al.} 2009, Phys. Rev. D, 79, 024017, \dodoi{10.1103/PhysRevD.79.024017}

\bibitem[{Dimmelmeier {et~al.}(2002)Dimmelmeier, Font, \&
  M{\"u}ller}]{Dimmelmeier02a}
Dimmelmeier, H., Font, J.~A., \& M{\"u}ller, E. 2002, Astron. Astrophys., 388,
  917

\bibitem[{Dudi {et~al.}(2022)Dudi, Dietrich, Rashti, Bruegmann, Steinhoff, \&
  Tichy}]{Dudi2021}
Dudi, R., Dietrich, T., Rashti, A., {et~al.} 2022, Phys. Rev. D, 105, 064050,
  \dodoi{10.1103/PhysRevD.105.064050}

\bibitem[{{Espino} {et~al.}(2023){Espino}, {Bozzola}, \&
  {Paschalidis}}]{Espino2023}
{Espino}, P.~L., {Bozzola}, G., \& {Paschalidis}, V. 2023, Phys. Rev. D, 107,
  104059, \dodoi{10.1103/PhysRevD.107.104059}

\bibitem[{{Etienne} {et~al.}(2015){Etienne}, {Paschalidis}, {Haas},
  {M{\"o}sta}, \& {Shapiro}}]{Etienne2015}
{Etienne}, Z.~B., {Paschalidis}, V., {Haas}, R., {M{\"o}sta}, P., \& {Shapiro},
  S.~L. 2015, Class. Quantum Grav., 32, 175009,
  \dodoi{10.1088/0264-9381/32/17/175009}

\bibitem[{{Faye} \& {Sch{\"a}fer}(2003)}]{Faye2003}
{Faye}, G., \& {Sch{\"a}fer}, G. 2003, Phys. Rev. D, 68, 084001,
  \dodoi{10.1103/PhysRevD.68.084001}

\bibitem[{{Fermi-LAT Collaboration}(2017)}]{FermiLat2017}
{Fermi-LAT Collaboration}. 2017, ArXiv e-prints.
\newblock \doarXiv{1710.05450}

\bibitem[{Fields {et~al.}(2025)Fields, Radice, \& Hammond}]{Fields2025}
Fields, J., Radice, D., \& Hammond, P. 2025.
\newblock \doarXiv{2507.18695}

\bibitem[{{Frieben} \& {Rezzolla}(2012)}]{Frieben2012}
{Frieben}, J., \& {Rezzolla}, L. 2012, Mon. Not. R. Astron. Soc., 427, 3406,
  \dodoi{10.1111/j.1365-2966.2012.22027.x}

\bibitem[{{Fujibayashi} {et~al.}(2023){Fujibayashi}, {Kiuchi}, {Wanajo},
  {Kyutoku}, {Sekiguchi}, \& {Shibata}}]{Fujibayashi2023}
{Fujibayashi}, S., {Kiuchi}, K., {Wanajo}, S., {et~al.} 2023, Astrophys. J.,
  942, 39, \dodoi{10.3847/1538-4357/ac9ce0}

\bibitem[{{Giacomazzo} {et~al.}(2009){Giacomazzo}, {Rezzolla}, \&
  {Baiotti}}]{Giacomazzo:2009mp}
{Giacomazzo}, B., {Rezzolla}, L., \& {Baiotti}, L. 2009, Mon. Not. R. Astron.
  Soc., 399, L164, \dodoi{10.1111/j.1745-3933.2009.00745.x}

\bibitem[{{Giacomazzo} {et~al.}(2011){Giacomazzo}, {Rezzolla}, \&
  {Baiotti}}]{Giacomazzo2011b}
---. 2011, Phys. Rev. D, 83, 044014, \dodoi{10.1103/PhysRevD.83.044014}

\bibitem[{{Giacomazzo} {et~al.}(2015){Giacomazzo}, {Zrake}, {Duffell},
  {MacFadyen}, \& {Perna}}]{Giacomazzo:2014a}
{Giacomazzo}, B., {Zrake}, J., {Duffell}, P.~C., {MacFadyen}, A.~I., \&
  {Perna}, R. 2015, Astrophys. J., 809, 39, \dodoi{10.1088/0004-637X/809/1/39}

\bibitem[{{Gottlieb} {et~al.}(2025){Gottlieb}, {Metzger}, {Foucart}, \&
  {Ramirez-Ruiz}}]{Gottlieb2024}
{Gottlieb}, O., {Metzger}, B.~D., {Foucart}, F., \& {Ramirez-Ruiz}, E. 2025,
  Astrophys. J., 984, 77, \dodoi{10.3847/1538-4357/adc577}

\bibitem[{Gottlieb {et~al.}(2023)}]{Gottlieb:2023a}
Gottlieb, O., {et~al.} 2023, Astrophys. J. Lett., 954, L21,
  \dodoi{10.3847/2041-8213/aceeff}

\bibitem[{Guti{\'e}rrez {et~al.}(2025)Guti{\'e}rrez, Cook, Radice, Bernuzzi,
  Fields, Hammond, Daszuta, Bandyopadhyay, \& Jacobi}]{Gutierrez2025}
Guti{\'e}rrez, E.~M., Cook, W., Radice, D., {et~al.} 2025.
\newblock \doarXiv{2506.18995}

\bibitem[{{Hajela} {et~al.}(2019){Hajela}, {Margutti}, {Alexander},
  {Kathirgamaraju}, {Baldeschi}, {Guidorzi}, {Giannios}, {Fong}, {Wu},
  {MacFadyen}, {Paggi}, {Berger}, {Blanchard}, {Chornock}, {Coppejans},
  {Cowperthwaite}, {Eftekhari}, {Gomez}, {Hosseinzadeh}, {Laskar}, {Metzger},
  {Nicholl}, {Paterson}, {Radice}, {Sironi}, {Terreran}, {Villar}, {Williams},
  {Xie}, \& {Zrake}}]{Hajela2019}
{Hajela}, A., {Margutti}, R., {Alexander}, K.~D., {et~al.} 2019, Astrophys. J.
  Lett., 886, L17, \dodoi{10.3847/2041-8213/ab5226}

\bibitem[{{Hajela} {et~al.}(2022){Hajela}, {Margutti}, {Bright}, {Alexander},
  {Metzger}, {Nedora}, {Kathirgamaraju}, {Margalit}, {Radice}, {Guidorzi},
  {Berger}, {MacFadyen}, {Giannios}, {Chornock}, {Heywood}, {Sironi},
  {Gottlieb}, {Coppejans}, {Laskar}, {Cendes}, {Duran}, {Eftekhari}, {Fong},
  {McDowell}, {Nicholl}, {Xie}, {Zrake}, {Bernuzzi}, {Broekgaarden},
  {Kilpatrick}, {Terreran}, {Villar}, {Blanchard}, {Gomez}, {Hosseinzadeh},
  {Matthews}, \& {Rastinejad}}]{Hajela2022_ag}
{Hajela}, A., {Margutti}, R., {Bright}, J.~S., {et~al.} 2022, Astrophys. J.
  Lett., 927, L17, \dodoi{10.3847/2041-8213/ac504a}

\bibitem[{Harten {et~al.}(1983)Harten, Lax, \& van Leer}]{Harten83}
Harten, A., Lax, P.~D., \& van Leer, B. 1983, SIAM Rev., 25, 35,
  \dodoi{10.1137/1025002}

\bibitem[{{Hessels} {et~al.}(2006){Hessels}, {Ransom}, {Stairs}, {Freire},
  {Kaspi}, \& {Camillo}}]{Hessels2006}
{Hessels}, J.~W., {Ransom}, S.~M., {Stairs}, I.~H., {et~al.} 2006, Science,
  311, 1901, \dodoi{10.1126/science.1123430}

\bibitem[{{Jiang} {et~al.}(2025){Jiang}, {Ng}, {Chabanov}, \&
  {Rezzolla}}]{Jiang2025}
{Jiang}, J.-L., {Ng}, H. H.-Y., {Chabanov}, M., \& {Rezzolla}, L. 2025, Phys.
  Rev. D, 111, 103043, \dodoi{10.1103/PhysRevD.111.103043}

\bibitem[{{Karakas} {et~al.}(2025){Karakas}, {Matur}, \&
  {Ruffert}}]{Karakas2024}
{Karakas}, B., {Matur}, R., \& {Ruffert}, M. 2025, Mon. Not. R. Astron. Soc.,
  \dodoi{10.1093/mnras/staf2009}

\bibitem[{Kawaguchi {et~al.}(2023)Kawaguchi, Fujibayashi, Domoto, Kiuchi,
  Shibata, \& Wanajo}]{Kawaguchi2023}
Kawaguchi, K., Fujibayashi, S., Domoto, N., {et~al.} 2023, arXiv preprint
  arXiv:2306.06961

\bibitem[{{Kawamura} {et~al.}(2016){Kawamura}, {Giacomazzo}, {Kastaun},
  {Ciolfi}, {Endrizzi}, {Baiotti}, \& {Perna}}]{Kawamura2016}
{Kawamura}, T., {Giacomazzo}, B., {Kastaun}, W., {et~al.} 2016, Phys. Rev. D,
  94, 064012, \dodoi{10.1103/PhysRevD.94.064012}

\bibitem[{Keppens {et~al.}(2021)Keppens, Teunissen, Xia, \&
  Porth}]{Keppens2021}
Keppens, R., Teunissen, J., Xia, C., \& Porth, O. 2021, Computers \&
  Mathematics with Applications, 81, 316

\bibitem[{Kiuchi(2024)}]{Kiuchi2024}
Kiuchi, K. 2024, arXiv preprint arXiv:2405.10081

\bibitem[{{Kiuchi} {et~al.}(2015){Kiuchi}, {Cerd{\'a}-Dur{\'a}n}, {Kyutoku},
  {Sekiguchi}, \& {Shibata}}]{Kiuchi2015a}
{Kiuchi}, K., {Cerd{\'a}-Dur{\'a}n}, P., {Kyutoku}, K., {Sekiguchi}, Y., \&
  {Shibata}, M. 2015, Phys. Rev. D, 92, 124034,
  \dodoi{10.1103/PhysRevD.92.124034}

\bibitem[{{Kiuchi} {et~al.}(2024){Kiuchi}, {Reboul-Salze}, {Shibata}, \&
  {Sekiguchi}}]{Kiuchi2023}
{Kiuchi}, K., {Reboul-Salze}, A., {Shibata}, M., \& {Sekiguchi}, Y. 2024,
  Nature Astronomy, 8, 298, \dodoi{10.1038/s41550-024-02194-y}

\bibitem[{{Lyman} {et~al.}(2018){Lyman}, {Lamb}, {Levan}, {Mandel}, {Tanvir},
  {Kobayashi}, {Gompertz}, {Hjorth}, {Fruchter}, {Kangas}, {Steeghs}, {Steele},
  {Cano}, {Copperwheat}, {Evans}, {Fynbo}, {Gall}, {Im}, {Izzo}, {Jakobsson},
  {Milvang-Jensen}, {O'Brien}, {Osborne}, {Palazzi}, {Perley}, {Pian},
  {Rosswog}, {Rowlinson}, {Schulze}, {Stanway}, {Sutton}, {Th{\"o}ne}, {de
  Ugarte Postigo}, {Watson}, {Wiersema}, \& {Wijers}}]{Lyman2018}
{Lyman}, J.~D., {Lamb}, G.~P., {Levan}, A.~J., {et~al.} 2018, Nature Astronomy,
  \dodoi{10.1038/s41550-018-0511-3}

\bibitem[{{Metzger}(2017)}]{Metzger2017}
{Metzger}, B.~D. 2017, Living Reviews in Relativity, 20, 3,
  \dodoi{10.1007/s41114-017-0006-z}

\bibitem[{{Most} {et~al.}(2020){Most}, {Jens Papenfort}, {Dexheimer},
  {Hanauske}, {Stoecker}, \& {Rezzolla}}]{Most2019c}
{Most}, E.~R., {Jens Papenfort}, L., {Dexheimer}, V., {et~al.} 2020, European
  Physical Journal A, 56, 59, \dodoi{10.1140/epja/s10050-020-00073-4}

\bibitem[{{Most} {et~al.}(2019){Most}, {Papenfort}, \& {Rezzolla}}]{Most2019b}
{Most}, E.~R., {Papenfort}, L.~J., \& {Rezzolla}, L. 2019, Mon. Not. R. Astron.
  Soc., 490, 3588, \dodoi{10.1093/mnras/stz2809}

\bibitem[{Most {et~al.}(2025)Most, Peterson, Scurto, Pais, \&
  Dexheimer}]{Most2025}
Most, E.~R., Peterson, J., Scurto, L., Pais, H., \& Dexheimer, V. 2025,
  Astrophys. J. Lett., 989, L29, \dodoi{10.3847/2041-8213/adf62d}

\bibitem[{{Most} \& {Quataert}(2023)}]{Most2023}
{Most}, E.~R., \& {Quataert}, E. 2023, Astrophys. J. Lett., 947, L15,
  \dodoi{10.3847/2041-8213/acca84}

\bibitem[{M\"osta {et~al.}(2020)M\"osta, Radice, Haas, Schnetter, \&
  Bernuzzi}]{Moesta2020}
M\"osta, P., Radice, D., Haas, R., Schnetter, E., \& Bernuzzi, S. 2020,
  Astrophys. J. Lett., 901, L37, \dodoi{10.3847/2041-8213/abb6ef}

\bibitem[{Musolino {et~al.}(2025)Musolino, Rezzolla, \& Most}]{Musolino2024b}
Musolino, C., Rezzolla, L., \& Most, E.~R. 2025, Astrophys. J. Lett., 984, L61,
  \dodoi{10.3847/2041-8213/adcd6d}

\bibitem[{Neuweiler {et~al.}(2025)}]{Neuweiler2025b}
Neuweiler, A., {et~al.} 2025.
\newblock \doarXiv{2510.14850}

\bibitem[{{Ng} {et~al.}(2021){Ng}, {Cheong}, {Lin}, \& {Li}}]{Ng2021}
{Ng}, H. H.-Y., {Cheong}, P. C.-K., {Lin}, L.-M., \& {Li}, T. G.~F. 2021, The
  Astrophysical Journal, 915, 108, \dodoi{10.3847/1538-4357/ac0141}

\bibitem[{{Ng} {et~al.}(2024){Ng}, {Jiang}, {Musolino}, {Ecker}, {Tootle}, \&
  {Rezzolla}}]{Ng2024b}
{Ng}, H. H.-Y., {Jiang}, J.-L., {Musolino}, C., {et~al.} 2024, Phys. Rev. D,
  109, 064061, \dodoi{10.1103/PhysRevD.109.064061}

\bibitem[{Ng {et~al.}(2025)Ng, Musolino, Tootle, \& Rezzolla}]{Ng2024c}
Ng, H. H.-Y., Musolino, C., Tootle, S.~D., \& Rezzolla, L. 2025, Astrophys. J.
  Lett., 985, L36, \dodoi{10.3847/2041-8213/add324}

\bibitem[{{Oechslin} {et~al.}(2007){Oechslin}, {Janka}, \&
  {Marek}}]{Oechslin07a}
{Oechslin}, R., {Janka}, H.-T., \& {Marek}, A. 2007, Astron. Astrophys., 467,
  395, \dodoi{10.1051/0004-6361:20066682}

\bibitem[{Olivares {et~al.}(2019)Olivares, Porth, Davelaar, Most, Fromm,
  Mizuno, Younsi, \& Rezzolla}]{Olivares2019}
Olivares, H., Porth, O., Davelaar, J., {et~al.} 2019, Astron. Astrophys., 629,
  A61, \dodoi{10.1051/0004-6361/201935559}

\bibitem[{{Ott} {et~al.}(2007){Ott}, {Dimmelmeier}, {Marek}, {Janka}, {Zink},
  {Hawke}, \& {Schnetter}}]{Ott07b}
{Ott}, C.~D., {Dimmelmeier}, H., {Marek}, A., {et~al.} 2007, Class. Quantum
  Grav., 24, 139, \dodoi{10.1088/0264-9381/24/12/S10}

\bibitem[{{Palenzuela} {et~al.}(2022){Palenzuela}, {Aguilera-Miret},
  {Carrasco}, {Ciolfi}, {Kalinani}, {Kastaun}, {Mi{\~n}ano}, \&
  {Vigan{\`o}}}]{Palenzuela_2022PRD}
{Palenzuela}, C., {Aguilera-Miret}, R., {Carrasco}, F., {et~al.} 2022, Phys.
  Rev. D, 106, 023013, \dodoi{10.1103/PhysRevD.106.023013}

\bibitem[{{Papenfort} {et~al.}(2021){Papenfort}, {Tootle}, {Grandcl{\'e}ment},
  {Most}, \& {Rezzolla}}]{Papenfort2021}
{Papenfort}, L.~J., {Tootle}, S.~D., {Grandcl{\'e}ment}, P., {Most}, E.~R., \&
  {Rezzolla}, L. 2021, arXiv e-prints, arXiv:2103.09911.
\newblock \doarXiv{2103.09911}

\bibitem[{{Paschalidis}(2017)}]{Paschalidis2016}
{Paschalidis}, V. 2017, Classical and Quantum Gravity, 34, 084002,
  \dodoi{10.1088/1361-6382/aa61ce}

\bibitem[{{Porth} {et~al.}(2017){Porth}, {Olivares}, {Mizuno}, {Younsi},
  {Rezzolla}, {Moscibrodzka}, {Falcke}, \& {Kramer}}]{Porth2017}
{Porth}, O., {Olivares}, H., {Mizuno}, Y., {et~al.} 2017, Computational
  Astrophysics and Cosmology, 4, 42, \dodoi{10.1186/s40668-017-0020-2}

\bibitem[{{Price} \& {Rosswog}(2006)}]{Price06}
{Price}, D.~J., \& {Rosswog}, S. 2006, Science, 312, 719,
  \dodoi{10.1126/science.1125201}

\bibitem[{{Radice} {et~al.}(2020){Radice}, {Bernuzzi}, \&
  {Perego}}]{Radice2020b}
{Radice}, D., {Bernuzzi}, S., \& {Perego}, A. 2020, Annual Review of Nuclear
  and Particle Science, 70, 95, \dodoi{10.1146/annurev-nucl-013120-114541}

\bibitem[{{Radice} {et~al.}(2016){Radice}, {Galeazzi}, {Lippuner}, {Roberts},
  {Ott}, \& {Rezzolla}}]{Radice2016}
{Radice}, D., {Galeazzi}, F., {Lippuner}, J., {et~al.} 2016, Mon. Not. R.
  Astron. Soc., 460, 3255, \dodoi{10.1093/mnras/stw1227}

\bibitem[{Rasio \& Shapiro(1999)}]{Rasio99__}
Rasio, F.~A., \& Shapiro, S.~L. 1999, Class. Quantum Grav., 16, R1

\bibitem[{{Rezzolla} {et~al.}(2011){Rezzolla}, {Giacomazzo}, {Baiotti},
  {Granot}, {Kouveliotou}, \& {Aloy}}]{Rezzolla:2011}
{Rezzolla}, L., {Giacomazzo}, B., {Baiotti}, L., {et~al.} 2011, Astrophys. J.
  Letters, 732, L6, \dodoi{10.1088/2041-8205/732/1/L6}

\bibitem[{{Rezzolla} \& {Zanotti}(2013)}]{Rezzolla_book:2013}
{Rezzolla}, L., \& {Zanotti}, O. 2013, {Relativistic Hydrodynamics} (Oxford
  University Press), \dodoi{10.1093/acprof:oso/9780198528906.001.0001}

\bibitem[{{Ruiz} {et~al.}(2018){Ruiz}, {Shapiro}, \& {Tsokaros}}]{Ruiz2018}
{Ruiz}, M., {Shapiro}, S.~L., \& {Tsokaros}, A. 2018, Phys. Rev. D, 98, 123017,
  \dodoi{10.1103/PhysRevD.98.123017}

\bibitem[{{Ruiz} {et~al.}(2020){Ruiz}, {Tsokaros}, \& {Shapiro}}]{Ruiz2020}
{Ruiz}, M., {Tsokaros}, A., \& {Shapiro}, S.~L. 2020, arXiv e-prints,
  arXiv:2001.09153.
\newblock \doarXiv{2001.09153}

\bibitem[{{Schnetter} {et~al.}(2004){Schnetter}, {Hawley}, \&
  {Hawke}}]{Schnetter-etal-03b}
{Schnetter}, E., {Hawley}, S.~H., \& {Hawke}, I. 2004, Class. Quantum Grav.,
  21, 1465, \dodoi{10.1088/0264-9381/21/6/014}

\bibitem[{{Siegel} {et~al.}(2013){Siegel}, {Ciolfi}, {Harte}, \&
  {Rezzolla}}]{Siegel2013}
{Siegel}, D.~M., {Ciolfi}, R., {Harte}, A.~I., \& {Rezzolla}, L. 2013, Phys.
  Rev. D R, 87, 121302, \dodoi{10.1103/PhysRevD.87.121302}

\bibitem[{Steinhoff {et~al.}(2021)Steinhoff, Hinderer, Dietrich, \&
  Foucart}]{Steinhoff2021}
Steinhoff, J., Hinderer, T., Dietrich, T., \& Foucart, F. 2021, Phys. Rev.
  Res., 3, 033129, \dodoi{10.1103/PhysRevResearch.3.033129}

\bibitem[{{The LIGO Scientific Collaboration} \& {The Virgo
  Collaboration}(2017)}]{Abbott2017}
{The LIGO Scientific Collaboration}, \& {The Virgo Collaboration}. 2017, Phys.
  Rev. Lett., 119, 161101, \dodoi{10.1103/PhysRevLett.119.161101}

\bibitem[{Togashi {et~al.}(2017)Togashi, Nakazato, Takehara, Yamamuro, Suzuki,
  \& Takano}]{Togashi2017}
Togashi, H., Nakazato, K., Takehara, Y., {et~al.} 2017, Nucl. Phys., A961, 78,
  \dodoi{10.1016/j.nuclphysa.2017.02.010}

\bibitem[{{Tootle} {et~al.}(2021){Tootle}, {Papenfort}, {Most}, \&
  {Rezzolla}}]{Tootle2021}
{Tootle}, S.~D., {Papenfort}, L.~J., {Most}, E.~R., \& {Rezzolla}, L. 2021,
  Astrophys. J. Lett., 922, L19, \dodoi{10.3847/2041-8213/ac350d}

\bibitem[{Yip \& Li(2025)}]{Yip2025}
Yip, A. K.~L., \& Li, T. G.~F. 2025.
\newblock \doarXiv{2509.10150}

\bibitem[{Yoshida(2012)}]{Yoshida2012b}
Yoshida, S. 2012, Phys. Rev. D, 86, 104055, \dodoi{10.1103/PhysRevD.86.104055}

\bibitem[{Zhu {et~al.}(2022)Zhu, Yang, Zhang, Gao, \& Yu}]{Zhu2021b}
Zhu, J.-P., Yang, Y.-P., Zhang, B., Gao, H., \& Yu, Y.-W. 2022, Astrophys. J.,
  938, 147, \dodoi{10.3847/1538-4357/ac8e60}

\bibitem[{{Zhu} {et~al.}(2020){Zhu}, {Li}, \& {Rezzolla}}]{Zhu2020}
{Zhu}, Z., {Li}, A., \& {Rezzolla}, L. 2020, Phys. Rev. D, 102, 084058,
  \dodoi{10.1103/PhysRevD.102.084058}

\end{thebibliography}

\end{document}